\documentclass[aps,pra,preprint,a4paper,groupedaddress,showpacs]{revtex4-1}
\usepackage{graphicx}
\usepackage{amsmath}
\usepackage{amssymb}
\usepackage{mathrsfs}
\usepackage{amsfonts}
\usepackage{dsfont}
\usepackage{dcolumn}
\usepackage{bm}
\usepackage{booktabs}
\usepackage{color}
\usepackage{xcolor}

\newcommand{\Vr}{\vec{r}}
\newcommand{\vrho}{\vec{\rho}}
\newcommand{\vk}{\vec{k}}
\newcommand{\vka}{\vec{\kappa}}
\newcommand{\mF}{\mathcal{F}}
\newcommand{\dd}{{d^{\,2}}}

\providecommand{\abs}[1]{\left|#1\right|}

\begin{document}
\title{Self-healing of Gaussian and Bessel beams: a critical comparison}
\author{Andrea Aiello$^{1,2}$}
\email{andrea.aiello@mpl.mpg.de}
\author{Girish S. Agarwal$^{3}$}
%
\affiliation{$^1$ Max Planck Institute for the Science of Light, G$\ddot{u}$nther-Scharowsky-Strasse 1/Bau24, 91058 Erlangen,
Germany}
\affiliation{$^2$Institute for Optics, Information and Photonics, University of Erlangen-Nuernberg, Staudtstrasse 7/B2, 91058 Erlangen, Germany}
\affiliation{$^3$ Department of Physics, Oklahoma State University, Stillwater, Oklahoma 74078, USA}
\date{\today}
\begin{abstract}
Contrarily to a common belief, any beam of light possesses to a some extent the ability to ``reconstruct itself'' after hitting an obstacle. The celebrated Arago spot phenomenon is nothing but a manifestation of this property. 
In this work we analyze the self-healing mechanism from both a mathematical and a physical point of view, eventually finding a new expression for the minimum reconstruction distance, which is valid for \emph{any} kind of beam, including Gaussian ones. Finally,  a witness function that quantify the self-reconstruction capability of a beam is proposed and tested.
The results presented here help clarifying the physics underlying self-healing mechanism in optical beams.

\end{abstract}
\maketitle

\section{Introduction}

In this paper we illustrate, with the help of  several examples,  the self-healing property exhibited by different types of optical beams. In particular, we analyze and compare the behavior of Gaussian, Bessel, Bessel-Gauss beams and of a pair of plane waves.
In recent years, the  remarkable capacity of a beam to reconstruct itself after encountering an obstacle (self-healing mechanism) attracted considerable interest and has been the subject of numerous theoretical and experimental investigations \cite{McGloin05,Jaregui05,Arlt01,Gar02,Fahrbach10,Fahrbach12,McLaren14}. In this context, particularly relevant is the theoretical work by Chu and coworkers \cite{Chu12,Chu14}. 

The aim of this work is to study further the self-reconstruction ability possessed by different kind of light beams, thus extending our previous work on this subject \cite{Aiello14}. For all the \emph{scalar} beams studied here, we model the obstacle as a soft-edge aperture with a Gaussian profile, in order to be able to produce formulas in closed form. In particular, after a careful analysis of the 
 self-healing mechanism  from a mathematical and physical point of view, we can find a new expression for the minimum reconstruction distance $z_\text{min}$. Then, a witness function that quantifies the similarity between the field of the unperturbed beam (namely, the beam that would propagate as if the obstacle were not present) and the field of the perturbed one (that is, the beam that propagates behind the obstruction), is proposed and tested. Finally, our findings are illustrated by means of a quantitative comparison between the self-healing properties of a Gauss beam and a  Bessel beam. The notation utilized throughout this paper is given in Appendix \ref{Not}.

\section{Self-healing and the eigenvalue problem}
\subsection{Basic definitions and formal developments}
Consider a scalar field  $f(x,y,z)$ propagating along the $z$-axis. An obstruction, characterized by a given amplitude transmission function $t_O (x,y)$, is placed in the plane $z=0$. The amplitude $f_O(x,y,0)$ of the obstructed field  in the plane $z=0$ can be written as 
\begin{align}\label{p10}
f_O(x,y,0) =  t_O(x,y)f(x,y,0).
\end{align}
According to the definition of angular spectrum, the amplitude $f_O(x,y,z)$ of the field transmitted at distance $z$ from the obstruction can be written as 
\begin{align}\label{p20}
f_O(x,y,z) =  \frac{1}{2 \pi} \iint\limits_{-\infty}^{\phantom{xx}\infty} F_O(k_x,k_y) \exp \left[i \left( x k_x + y k_y + z k_z\right) \right] \, dk_x dk_y,
\end{align}
 where $k_z = \left(k^2 - \kappa^2 \right)^{1/2}$, with $\kappa^2 = k_x^2 + k_y^2$ and $F_O(k_x,k_y) = \mF\left[f(x,y,0) t_O(x,y) \right](k_x,k_y)$, namely
\begin{align}\label{p30}
F_O(k_x,k_y) = & \; \frac{1}{2 \pi} \iint\limits_{-\infty}^{\phantom{xx}\infty} f(x,y,0) t_O(x,y) \exp \left[- i \left( x k_x + y k_y\right) \right] \, dx dy \nonumber \\
 = & \; \frac{1}{2 \pi} \iint\limits_{-\infty}^{\phantom{xx}\infty} 
\left[ \iint\limits_{-\infty}^{\phantom{xx}\infty} F(\vka{\,'})\,e^{i \vrho \cdot \vka{\,'}} \frac{{d^{\,2}}  \kappa'}{2 \pi} \right]
\! \left[ \iint\limits_{-\infty}^{\phantom{xx}\infty} T_O(\vka{\,''})\,e^{i \vrho \cdot \vka{\,''}} \frac{{d^{\,2}}  \kappa''}{2 \pi} \right]\,e^{-i \vrho \cdot \vka}
 \, dx dy \nonumber \\
= & \; \frac{1}{2 \pi} \iint\limits_{-\infty}^{\phantom{xx}\infty} {{d^{\,2}}  \kappa'} \left\{ F(\vka{\,'}) \iint\limits_{-\infty}^{\phantom{xx}\infty} {{d^{\,2}}  \kappa''}  \left[ T_O(\vka{\,''})\iint\limits_{-\infty}^{\phantom{xx}\infty}e^{i \vrho \cdot (\vka{\,'}+\vka{\,''}-\vka)} \, \frac{dx dy}{(2 \pi)^2} \right] \right\} \nonumber \\
= & \; \frac{1}{2 \pi} \iint\limits_{-\infty}^{\phantom{xx}\infty}   F(\vka{\,'})T_O(\vka - \vka{\,'}) {{d^{\,2}}  \kappa'},
\end{align}
where $\dd \kappa = d \kappa_x d \kappa_y$, $\dd \kappa' = d \kappa_x' d \kappa_y'$, et cetera. The final expression for $f_O(x,y,z)$ is thus given by
\begin{align}\label{p35}
f_O(x,y,z) =   \frac{1}{(2 \pi)^2} \iint\limits_{-\infty}^{\phantom{xx}\infty} \exp \left(i \vrho \cdot \vka\right) 
\exp \left(i z k_z \right)
\left[\iint\limits_{-\infty}^{\phantom{xx}\infty} T_O(\vka - \vka{\,'}) F(\vka{\,'}) \, \dd \kappa' \right] \dd \kappa.
\end{align}
Given the function $t_O(x,y)$, one  can always define the transmission function $t_A(x,y)$ of an aperture \emph{complementary} to the obstruction via the relation (Babinet principle)
\begin{align}\label{p182}
t_A(x,y) + t_O(x,y) =1.
\end{align}
By definition, both $t_A(x,y)$ and $t_O(x,y)$ are non-negative real-valued functions.  Using this equation into Eq. \eqref{p10} yields to 
\begin{align}\label{p10b}
f_O(x,y,0) =  & \; \left[ 1 -t_A(x,y) \right]f(x,y,0) \nonumber \\
=  & \; f(x,y,0)  - t_A(x,y)f(x,y,0)\nonumber \\
\equiv   & \; f(x,y,0)  - f_A(x,y,0).
\end{align}
Substituting Eq. \eqref{p10b} into Eq. \eqref{p35} one obtains
\begin{align}\label{p35b}
f_O(x,y,z) =  & \; \frac{1}{(2 \pi)^2} \iint\limits_{-\infty}^{\phantom{xx}\infty} \exp \left(i \vrho \cdot \vka\right) 
\exp \left(i z k_z \right)
\Biggl[ \Biggr. \nonumber \\
 & \; \times \Biggl. F(\vka) - 
\iint\limits_{-\infty}^{\phantom{xx}\infty} T_A(\vka - \vka{\,'}) F(\vka{\,'}) \, \dd \kappa' \Biggr] \dd \kappa \nonumber \\
\equiv & \; f(x,y,z) - f_A(x,y,z),
\end{align}
where $T_A(\vka) =  \mF[t_A](k_x,k_y)$ and the definition of the ``transmitted'' function $ f_A(x,y,z)$ is obvious. Taking the absolute value squared of both sides of either Eq. \eqref{p10b} or  Eq. \eqref{p35b} (the integrals are anyway independent from $z$)  and integrating over the \emph{whole} $xy$-plane, we obtain
\begin{align}\label{Lambda}
\overline{I}(f_O)= & \;  \overline{I}(f - f_A) \nonumber \\
= & \; \overline{I}(f) + \overline{I}(f_A) - 2 \operatorname{Re} \iint\limits_{-\infty}^{\phantom{xx}\infty} f^*(x,y,0) f_A(x,y,0)  \, dx dy,
\end{align}
where the average beam intensity $\overline{I}(h)$ is defined as
\begin{align}\label{k8}
\overline{I}(h)  = \iint\limits_{-\infty}^{\phantom{xx}\infty} h^*(x,y,z) h(x,y,z)  \, dx dy. 
\end{align}
 Equation \eqref{Lambda} already shows that \emph{perfect} self-healing is impossible even in principle because the intensity of the transmitted field is unavoidably reduced unless $\overline{I}(f_A) =0$. 
\subsection{Defining self-healing}
An optical beam is dubbed ``self-reconstructing'' when  has the ability to
recover its initial amplitude or intensity profile after interaction with an obstacle. This means that the field of a self-reconstructing beam is expected to obey the law
\begin{align}\label{p40}
 f_O(x,y,z) \approx   \lambda_0 f \! \left(x,y, z \right), 
\end{align}
for $z \geq z_0$, where $z_0$ denotes the so-called \emph{minimum reconstruction distance} and the scaling factor $ \lambda_0=  \left[\overline{I}(f_O)/ \overline{I}(f)\right]^{1/2}
$ accounts for the average intensity reduction caused by the interaction with the obstruction.  
The left side of Eq. \eqref{p40} is given by Eq. \eqref{p35}, while the right side can be written as
\begin{align}\label{p50}
\lambda_0 f \! \left(x,y,z\right) =  & \; \frac{\lambda_0}{2 \pi} \iint\limits_{-\infty}^{\phantom{xx}\infty} F(k_x,k_y) \exp \left[i \left( x k_x + y k_y + z k_z \right) \right] \, dk_x dk_y \nonumber \\
=  & \; \frac{1}{2 \pi} \iint\limits_{-\infty}^{\phantom{xx}\infty}
\exp \left(i \vrho \cdot \vka\right) 
\exp \left(i z k_z \right) \Bigl[ \lambda_0 F(\vka)  \Bigr] \, \dd \kappa. 
\end{align}
Substituting Eqs. \eqref{p35} and \eqref{p50} into \eqref{p40}, we obtain the following simple  relation: 
\begin{align}\label{p150}
F_O(\vka)\approx \lambda_0  F(\vka) ,
\end{align}
where  Eq. \eqref{p30} has been used.

It is crucial to notice that Eq. \eqref{p150} \emph{does not} contain the variable $z$ while, conversely,  the relation \eqref{p40} is supposed to be true only for  $z \geq z_0$. The latter requirement cannot be ignored because by definition Eq. \eqref{p40} cannot be satisfied at $z=0$ where, instead, Eq. \eqref{p10b} must be fulfilled.
 Hence,  we are apparently faced with a paradox here. In fact, Eq. \eqref{p150}  constitutes  more a statement 
about the obstruction rather than the field. This can be seen by using 
 Eq. \eqref{p50} to rewrite \eqref{p150} in the more suggestive form
\begin{align}\label{p180}
 \frac{1}{2 \pi} \iint\limits_{-\infty}^{\phantom{xx}\infty} T_O(\vka - \vka{\,'}) F(\vka{\,'}) \, \dd \kappa' 
\approx \lambda_0   F(\vka)  . 
\end{align}
If we replace the approximation symbol ``$\approx$'' with the equality one ``$=$'' in Eq. \eqref{p180}, then the latter  takes the form of a \emph{homogeneous  Fredholm integral equation of the second kind} \cite{Volterra} with the unknown function $F(\vka)$, which has to be 
an eigenfunction, associated with the eigenvalue $\lambda_0$, of the integral kernel $ T_O(\vka - \vka{\,'})$ describing  the obstruction. This means that the requirement \eqref{p40} is indeed too much restrictive because it can be satisfied only by those beams whose angular spectrum (the eigenfunction) is unaffected by the interaction with the obstruction, apart from a trivial proportionality factor (the eigenvalue), as shown in Eq. \eqref{p180}.
%
%
%
\begin{figure}[h!]
\centerline{\includegraphics[scale=3,clip=false,width=.4\columnwidth,trim = 0 0 0 0]{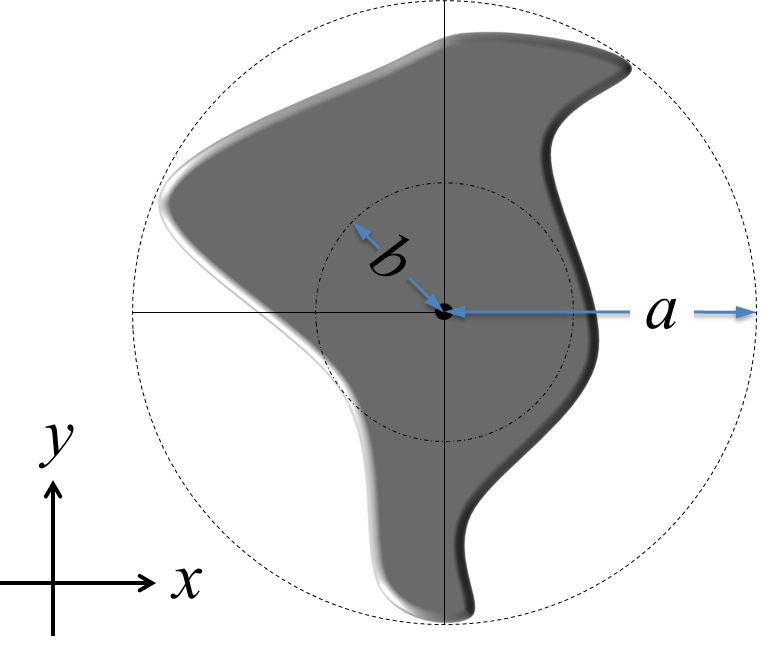}}
\caption{\label{fig1}
The orthogonal projection of the obstruction on the $xy$-plane is represented by the dark-gray area. This region is circumscribed by the dashed circle of radius $a$ (circumradius) and it inscribes the dot-dashed circle of radius $b$ (inradius).
Both circles are centered along the $z$-axis of the beam at $x=0=y$.}
\end{figure}
%
%

\subsection{Redefining the minimum reconstruction distance}
Key to the apparent paradox illustrated above, is the existence of the minimum reconstruction distance $z_0$ after which a self-reconstructing beam restores its initial profile. For a single plane wave   $\exp(i \vk \cdot \Vr)$ with wave vector $\vk = k \left(\hat{x} \sin \theta \cos \phi + \hat{y} \sin \theta \sin \phi + \hat{z} \cos \theta \right)$, this parameter can be straightforwardly determined in the context of either geometrical and wave optics \cite{Aiello14}.  Consider an obstruction whose orthogonal projection on the $xy$-plane occupies an area $O$,  and let $a$ be the \emph{radius} (circumradius) of the circle inside which such projection can be inscribed and centered on the axis of the beam (see Fig. \ref{fig1} above). 
Then $z_0$ can be estimated as 
\begin{align}\label{z10}
z_0 \propto \frac{a}{\tan \theta} ,
\end{align}
where the proportionality factor essentially depends on the shape of the obstruction. However, a beam of light can always be thought as a bunch of plane waves whose density in the $k$-space is determined by  the absolute value squared $\abs{F(k_x,k_y)}^2$ of the angular spectrum of the field $f(x,y,z)$ representing the beam. Moreover, given the wave vector  $\vk = k \left(\hat{x} \sin \theta \cos \phi + \hat{y} \sin \theta \sin \phi + \hat{z} \cos \theta \right) = \hat{x} k_x + \hat{y} k_y  + \hat{z}  k_z$, it is evidently possible to rewrite 
\begin{align}\label{z20}
\frac{1}{\tan \theta}   = & \;  \frac{k_z}{\left( k_x^2 + k_y^2 \right)^{1/2}} \nonumber \\
  = & \;  \frac{\left( k^2 -  k_x^2 - k_y^2 \right)^{1/2}}{\left( k_x^2 + k_y^2 \right)^{1/2}},
\end{align}
provided that $ k_x^2 + k_y^2 \leq k^2$. This condition is necessary to maintain  $k_z$ real-valued and it limits the applicability of the equation above to beams whose angular spectrum does not contain evanescent waves \cite{MandelBook}.
From Eq. \eqref{z20} it follows that we can regard the expression of $z_0$ given in Eq. \eqref{z10} as  a function of $\kappa = (k_x^2 + k_y^2 )^{1/2}$ in the $k$-space, namely
\begin{align}\label{z30}
z_0 \; \to \; a \, Z(\kappa) =   a \, \frac{\left( k^2 - \kappa^2 \right)^{1/2}}{\kappa}.
\end{align}
On account of the fact that the transverse wave vector $\vka$ has a  density distribution function $\abs{F(k_x,k_y)}^2$, 
 we  (arbitrarily) \emph{define}  the minimum reconstruction distance $z_0$ as the \emph{expected value} of the function $a \, Z(\kappa)$, namely
\begin{align}\label{z40}
\frac{z_0}{a} = \langle Z(\kappa) \rangle =  \frac{\phantom{ix} \displaystyle{\iint\limits_{k_x^2 + k_y^2 \leq k^2} \frac{\left( k^2 - \kappa^2 \right)^{1/2}}{\kappa} \abs{F(\vka)}^2\, \dd \kappa}}
{\displaystyle{\iint\limits_{k_x^2 + k_y^2 \leq k^2} \abs{F(\vka)}^2 \, \dd \kappa }},
\end{align}
where both integrals are limited to the disk of equation $k_x^2 + k_y^2 \leq k^2$.
It is convenient to express the integrals in Eq.  \eqref{z40} in cylindrical coordinates $k_x = \kappa \cos \varphi, \, k_y = \kappa \sin \varphi$ to obtain
\begin{align}\label{z50}
\frac{z_0}{a}  = & \; \frac{\displaystyle{\int\limits_{0}^{\phantom{xx}2 \pi }d \varphi \int\limits_{0}^{\phantom{xx} 1} \, ds \left( 1 - s^2 \right)^{1/2} \mathcal{F}(s,\varphi) }}
{\displaystyle{\int\limits_{0}^{\phantom{xx}2 \pi }d \varphi \int\limits_{0}^{\phantom{xx} 1 } d s \, s \, \mathcal{F}(s,\varphi)}} \nonumber \\
 = & \; \frac{\displaystyle{\int\limits_{0}^{\phantom{xx}2 \pi }d \varphi \int\limits_{0}^{\phantom{xx} \pi/2} \, d \theta \cos^2 \! \theta \,  \mathcal{F}(\sin \theta,\varphi) }}
{\displaystyle{\int\limits_{0}^{\phantom{xx}2 \pi }d \varphi \int\limits_{0}^{\phantom{xx} \pi/2 } d \theta \, \sin \theta \cos \theta \, \mathcal{F}(\sin \theta,\varphi)}},
\end{align}
where we have defined $s = \kappa/k = \sin \theta$ and $\mathcal{F}(s,\varphi) = \abs{F(\kappa \cos \varphi,\kappa \sin \varphi)}^2$.

In the remainder of this section, we check the validity of Eq. \eqref{z50} for the cases of \emph{1.} a \emph{fundamental Gaussian beam} and \emph{2.} a \emph{Bessel-Gauss beam}; both obstructed by a soft-edge Gaussian aperture. For the sake of clarity in the following examples we shall restrict our attention to the paraxial regime of propagation.

\subsubsection{Fundamental Gaussian beam}\label{Gauss}

\begin{figure}[h!]
\centerline{\includegraphics[scale=3,clip=false,width=1\columnwidth,trim = 0 0 0 0]{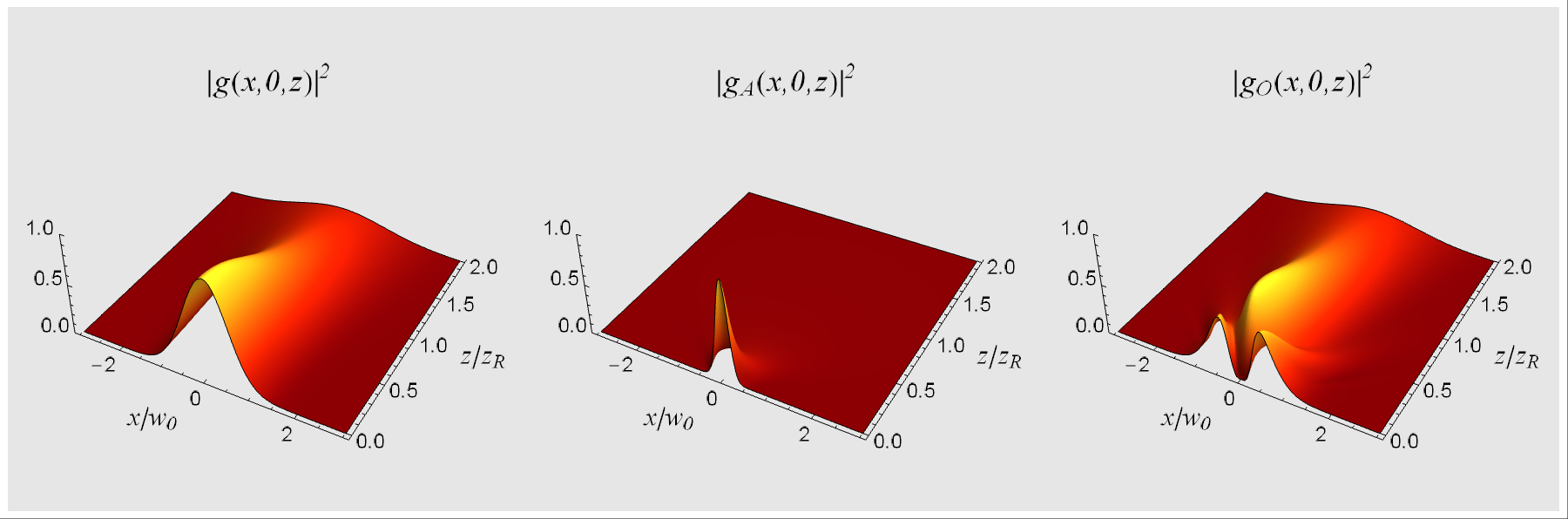}}
 \caption{ \label{fig2}
Plots of the intensity distributions (absolute value squared) evaluated at $y=0$, of (left to right): the incident field $g(x,0,z)$, the ``virtual'' field transmitted by the aperture complementary to the obstruction $g_A(x,0,z)$, and the field transmitted behind the obstacle $g_O(x,0,z)$. The plots are generated for a Gaussian beam with angular aperture  $\theta_0 = 2/(k w_0) = \pi/12 \sim 15^\circ$ and a soft-edge Gaussian obstruction with full width $a/w_0 = 0.2$ and $x_0 = 0 =y_0$. 
 At $z/z_R = 2$ the intensity profiles of the fields $g(x,0,z)$ and $g_O(x,0,z)$ appear  very similar.}
\end{figure}

\begin{figure}[h!]
\centerline{\includegraphics[scale=3,clip=false,width=.5\columnwidth,trim = 0 0 0 -100]{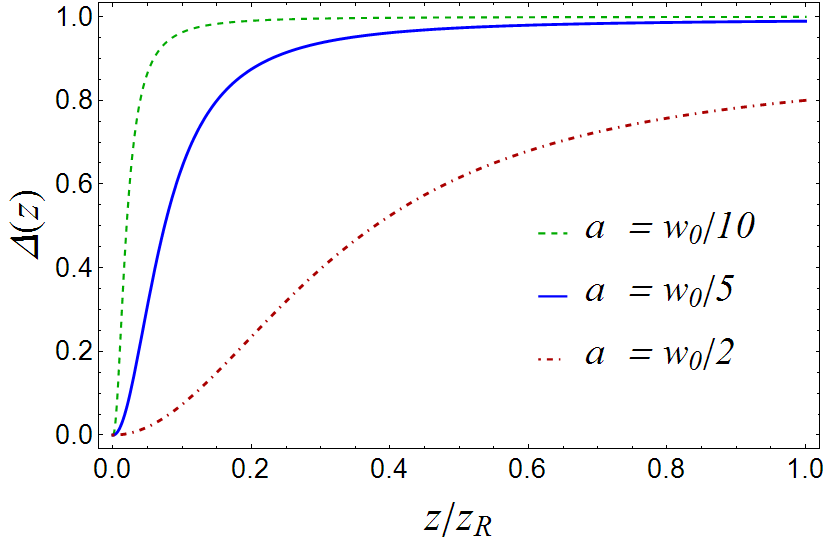}}
\caption{\label{fig3}
Plots of the the normalized difference $\Delta(z)$ of the \emph{on-axis} field intensities $I(0,0,z) = \abs{g(0,0,z)}^2$ and $I_A(0,0,z) = \abs{g_A(0,0,z)}^2$. For long propagation distances $z \gg z_R$ the asymptotic value is $\Delta(\infty) = 1 - (a_R/z_R)^2= (1+4 \alpha^2)/(1 + 2 \alpha^2)^2$, where $\alpha \equiv a/w_0$.}
\end{figure}

\begin{figure}[h!]
\centerline{\includegraphics[scale=3,clip=false,width=1\columnwidth,trim = 0 100 0 100]{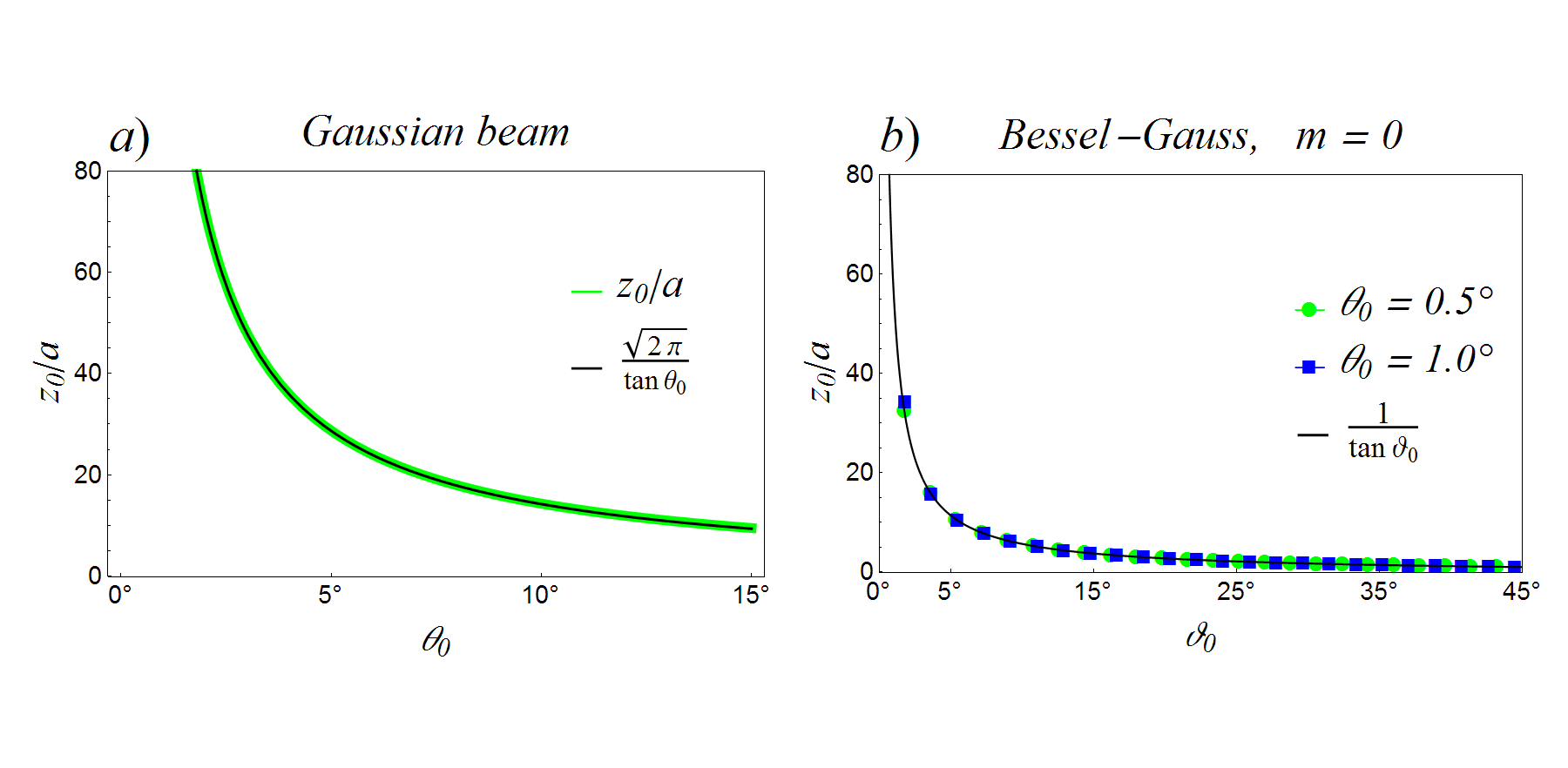}}
\caption{\label{fig4}
Plots of the minimum reconstruction distance $z_0/a$ as given in Eq. \eqref{z50}, evaluated for  \emph{a}) a Gaussian beam  and \emph{b})  a zeroth-order $(m=0)$ Bessel-Gauss beam. \emph{a}) Plot of  Eq. \eqref{z180} (green line) compared with the modified geometrical-optics value (black line). \emph{b}) Scatter plots of numerically integrated Eq. \eqref{z50} with $\mathcal{F}(s,\varphi)$ given by Eq. \eqref{z220} for different angular spreads $\theta_0$ of the Gaussian  envelope.  Green circles: $\theta_0 = 0.5^\circ$;  blue squares: $\theta_0 = 1.0^\circ$. The black line represents the ``one-over-tangent'' geometrical-optics  prediction  and is presented for comparison. }
\end{figure}

Consider the transmission of a fundamental Gaussian beam of waist $w_0$ across a soft-edge Gaussian obstacle of full width $2a$ located along the axis of the beam at $z=0$. The obstruction is described by the transmission function
\begin{align}\label{pz100}
t_O(x,y) = 1 - \exp\left(-  \frac{\abs{ \vrho -\vrho_0}^2}{2 a^2}\right),
\end{align}
where $\vrho_0 = \hat{x} x_0 + \hat{y} y_0$ represents the  displacement of the obstacle with respect to the beam propagation axis.
A straightforward calculation gives 
\begin{align}\label{z110}
T_O(k_x,k_y) = 2 \pi \delta \left( \vka\right)- a^2 \exp \left( - i \,\vka \cdot \vrho_0\right) \exp\left[  - \frac{ a^2}{2} \left( k_x^2 + k_y^2\right)\right].
\end{align}
The field describing the Gaussian beam can be written as $f(x,y,z) = \exp\left( i k z \right)g(x,y,z)$, with $g(x,y,z)$ being the fundamental solution of the paraxial wave equation
\begin{align}\label{z120}
g(x,y,z) = \frac{1}{z - i z_R} \exp\left[  i \frac{k}{2} \left(\frac{ x^2 + y^2}{z - i z_R}\right)\right],
\end{align}
and $z_R = k w_0^2/2$ denotes the Rayleigh range. The Fourier transform at $z=0$ of this field can be easily calculated and the result is
\begin{align}\label{z130}
G(k_x,k_y) = \frac{i}{k} \exp\left[  - \frac{z_R}{2 k} \left( k_x^2 + k_y^2\right)\right].
\end{align}
From Eqs. (\ref{p30},\ref{z120}-\ref{z130}) and a straightforward Gaussian integration, we obtain
\begin{align}\label{z140}
G_A(k_x,k_y) = \frac{i a^2}{a^2 k + z_R} \exp\left[  -\frac{k}{2} \frac{ \abs{\vrho_0}^2  }{a^2 k + z_R} - i \vka \cdot \vrho_0\frac{ z_R  }{a^2 k + z_R}\right] \exp \left( -\frac{a^2}{2} \frac{k_x^2 + k_y^2}{a^2 k + z_R}\right),
\end{align}
with, by definition, $G_O(k_x,k_y) = G(k_x,k_y) - G_A(k_x,k_y)$.
Using this result into Eq. \eqref{p20} yields to the following expression for the beam transmitted beyond the obstacle:
\begin{align}\label{z150}
g_A(x,y,z) = \frac{a_R}{z_R}\frac{1}{z - i a_R} \exp\left[  i \frac{k}{2} \left(\frac{ x^2 + y^2}{z - i a_R}\right)\right],
\end{align}
where, for the sake of clarity, we have chosen $\vrho_0 = \vec{0}$ and we have defined the modified Rayleigh range $a_R$ of the beam $g_A(x,y,z)$ transmitted by the aperture complementary to the obstruction, as
\begin{align}\label{z160}
a_R = \frac{\displaystyle{z_R}}{\displaystyle{1 + \frac{z_R}{k a^2}}} \leq z_R.
\end{align}
The self-healing capability of a fundamental Gaussian beam is vividly illustrated in Fig. \ref{fig2}.
%
%
%
%
\\
A close inspection of this figure together with Eqs. (\ref{z150}-\ref{z160}) reveals how the mechanism underlying  the self-reconstruction process works. From Eq. \eqref{z160} it follows that $a_R  \leq z_R$. Ergo, the ``virtual'' field $g_A(x,y,z)$ transmitted by the complementary aperture spreads in the $xy$-plane, while propagating along the $z$-axis, much more rapidly than the unperturbed field $g(x,y,z)$. Therefore, for $z \gtrsim 2$ the intensity profile of the obstructed beam almost coincides with the profile of the unperturbed one. This process is depicted  in Fig. \ref{fig3} where the normalized difference $\Delta(z)$ of the \emph{on-axis} field intensities $I(0,0,z) = \abs{g(0,0,z)}^2$ and $I_A(0,0,z) = \abs{g_A(0,0,z)}^2$ is plotted as function of $z/z_R$:
\begin{align}\label{z170}
\Delta(z) = \frac{I(0,0,z) - I_A(0,0,z)}{I(0,0,z)}.
\end{align}
%
%
%
%
%
From Eq. \eqref{z130} it follows that  $\mathcal{F}(s,\varphi) =\abs{G( \kappa \cos \varphi, \kappa \sin \varphi)}^2 = \exp\left( -2 s^2/\theta_0^{\, 2}\right)$, where $\theta_0 = 2/(k w_0)$ denotes the so-called \emph{angular spread} of the  Gaussian beam \cite{MandelBook}. Using this result in  Eq. \eqref{z50} yields to
\begin{align}\label{z180}
\frac{z_0}{a} = & \; \frac{\pi}{2 \theta_0^{\, 2}} \, \frac{\text{I}_0(1/\theta_0^{\, 2}) + \text{I}_1(1/\theta_0^{\, 2})}{\sinh \left( 1/\theta_0^{\, 2} \right)}  \\ \nonumber
\approx & \; \frac{\left( 2\pi \right)^{1/2}}{\theta_0} \frac{1}{1 - \exp \left(-2/\theta_0^{\, 2} \right)} , \qquad \quad \left( \theta_0 \ll 1 \right),
\end{align}
where $\text{I}_\nu(z)$ denotes the \emph{modified Bessel function of the first kind} of order $\nu$ \cite{G&R}. A plot of ${z_0}/{a}$ is given in Fig. \ref{fig4} \emph{a}). Since for $\theta_0 \ll 1$ it has $\theta_0 \approx \tan \theta_0$, then in the paraxial regime of propagation
\begin{align}\label{z190}
\frac{z_0}{a} 
\approx  \frac{\left( 2\pi \right)^{1/2}}{\tan \theta_0} , \qquad \quad \left( \theta_0 \ll 1 \right),
\end{align}
which is consistent with the expected geometrical optics result.

%
%
%
%

\subsubsection{Bessel-Gauss beam}

Consider a Bessel-Gauss beam transmitted across the same soft-edge Gaussian obstruction as above. The field describing such beam can be written in $z=0$ as 
\begin{align}\label{z200}
f(x,y,0) = \left( \pm 1 \right)^m \exp \left( \pm i m \phi \right) J_m(\kappa_0 \rho) \exp\left(- \frac{\rho^2}{w_0^2} \right), \qquad (m=0,1,2, \cdots),
\end{align}
where $\rho = (x^2 + y^2)^{1/2}$, $\kappa_0 = k \sin \vartheta_0$ and $J_\nu(z)$ denotes the  Bessel function of the first kind of order $\nu$ \cite{G&R}. The two key parameters characterizing the beam are the  aperture $\vartheta_0$ of the Bessel cone in $k$-space, and the waist $w_0$ (or, equivalently, the angular spread $\theta_0 = 2/(k w_0)$) of the Gaussian envelope.
 The Fourier transform at $z=0$ of this field is given, for $m \geq 0$, by the following expression:
\begin{align}\label{z210}
F(k_x,k_y) = \frac{w_0^2}{2 \, i^m} \exp \left( \pm i m \varphi \right) \text{I}_m \left( \frac{\kappa \, \kappa_0}{2/w_0^2}\right) \exp\left(  - \frac{\kappa^2 + \kappa_0^2}{4/w_0^2} \right).
\end{align}
Taking the absolute value squared of the function above yields to 
\begin{align}\label{z220}
\mathcal{F}(s,\varphi) = I_m^{\,2} \left( s \, \frac{2 \sin \vartheta_0}{\theta_0^{\,2}}\right) 
\exp \left[-  \frac{2 }{\theta_0^{\, 2}} \left(s^2 + \sin^2 \vartheta_0 \right)\right].
\end{align}
In this case the integrals in   Eq. \eqref{z50} cannot be calculated analytically and a numerical evaluation of the latter is necessary. The resulting values of $z_0/a$ are portrayed in Fig. \ref{fig4} \emph{b}) as functions of the  aperture  $\vartheta_0$, non necessarily paraxial, of the Bessel cone.

For $\theta_0 \ll 1$ the Bessel-Gauss beam reduces to an ordinary Bessel beam.  In such a limit we recover the expected geometrical optics result:
\begin{align}\label{z230}
\frac{z_0}{a} \approx  \frac{1}{\tan \vartheta_0} , \qquad \quad \left( \theta_0 \ll 1 \right).
\end{align}

The two examples shown in this section show that Eq. \eqref{z40} furnishes an appropriate measure of the minimum reconstruction distance $z_0$.

\section{Resolution of the paradox}

In this section we seek a solution for the paradox outlined earlier. The conundrum may be stated as follows: How is it possible to obtain the simultaneous validity of both Eq. \eqref{p10} and Eq. \eqref{p40}? In mathematical terms, the problem amounts to understand if and how it is possible achieve both 
\begin{align}\label{q10}
 f_O(x,y,0) = t_O(x,y)f(x,y,0) \qquad \text{AND} \qquad  f_O(x,y,z) \approx   \lambda_0 f \! \left(x,y, z \right) \qquad \forall z\geq z_0 .
\end{align}
As it will be clear soon, the solution of this problem is closely connected to the question raised by Chu\&Wen \cite{Chu14} about how to quantify the similarity between the two functions $f_O(x,y,z)$ and $f(x,y,z)$ in order to describe the self-reconstruction ability of a  beam. The reader is addressed to Appendix \ref{CW} for a shortly review of the Chu\&Wen approach.

\subsection{Similarity as relative distance}

Let us begin our discussion about similarity between functions by illustrating a simple, prototypical example of a self-reconstructing beam. 
Consider a ``skeleton'' Bessel beam made of two plane waves only, with wave vectors lying on the $xz$-plane and forming the angles $\vartheta_0$ and $-\vartheta_0$, respectively, with the $z$-axis. Let $f(x,y,z)$ be the complex-valued scalar field representing such a ``beam'' for $z \leq 0$:
\begin{align}\label{k50}
f(x,y,z) = \frac{1}{2} \exp \left( i k z \cos \vartheta_0\right)\left[ \exp \left( i k x \sin \vartheta_0\right)+ \exp \left( -i k x \sin \vartheta_0\right) \right].
\end{align}
 This beam hits at $z=0$ a semitransparent obstacle described by the Gaussian transmission function \eqref{pz100} that here we rewrite as
\begin{align}\label{k60}
t_O(x,y) = 1 - \exp \left[ - \frac{\left(x^2 -x_0 \right)^2 + \left( y- y_0 \right)^2}{2 a^2} \right],
\end{align}
with $a > 0$. The field $f_O(x,y,z)$ transmitted beyond the obstacle at $z>0$ can be straightforwardly  calculated and the expression is
\begin{align}\label{k70}
f_O(x,y,z) = & \; \frac{1}{1 + i{z}/({k a^2})} \exp \left[ - \frac{1}{2 a^2} \frac{\left(x - x_0 \right)^2 + \left( y - y_0 \right)^2}{1 + i{z}/({k a^2})}\right]
\exp \left[ -i\frac{k \sin^2 \vartheta_0}{2} \frac{z}{1 + i{z}/({k a^2})} \right] \nonumber \\
& \; \times \cos \left[ k \sin \vartheta_0 \, \frac{x + i x_0{z}/({k a^2})}{1 + i{z}/({k a^2})} \right].
\end{align}
It is easy to verify that, as expected,
\begin{align}\label{k80}
\lim_{a \to \infty }f_O(x,y,0) = f(x,y,0).
\end{align}
The  ``virtual'' field $f_A(x,y,z)$ transmitted by the aperture complementary to the obstacle can be obtained from the relation
\begin{align}\label{k90}
 f(x,y,z) = f_O(x,y,z) + f_A(x,y,z).
\end{align}
The absolute value squared of the three fields $f, f_A$ and $f_O$ is shown in Fig. \ref{fig5} below.
%
%
\begin{figure}[h]
\centerline{\includegraphics[scale=3,clip=false,width=1\columnwidth,trim = 0 0 0 0]{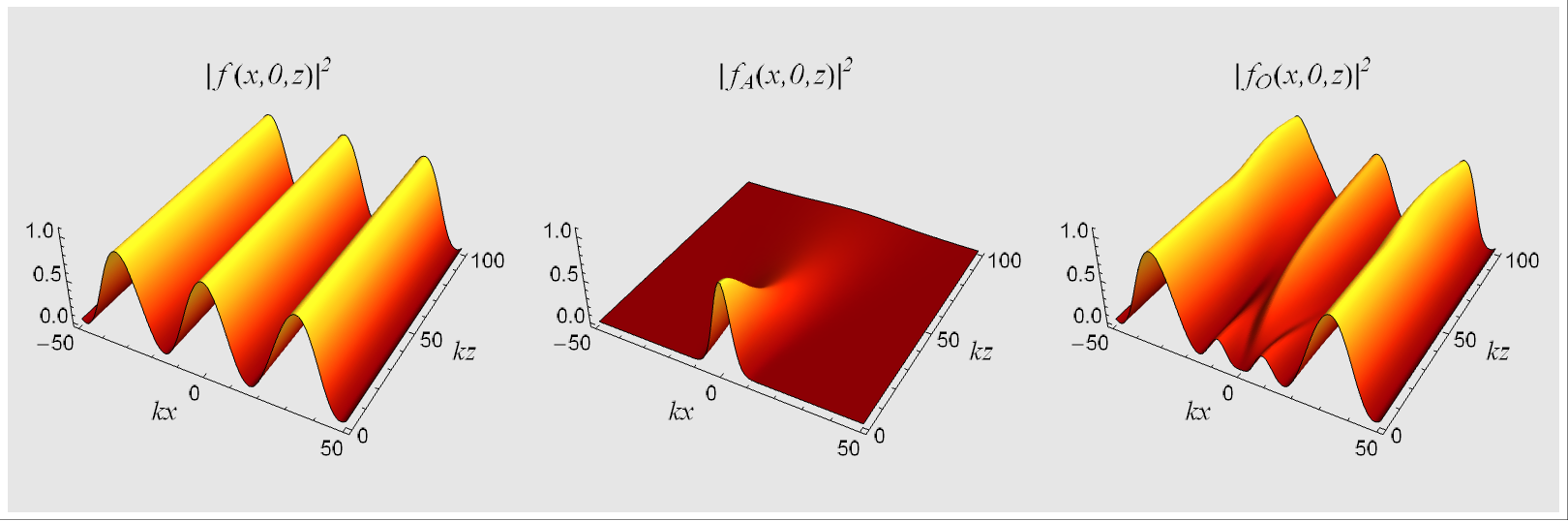}}
\caption{\label{fig5}
Plots of the intensity distributions (absolute value squared) evaluated at $y=0$, of (left to right): the original field $f(x,y,z)$, the ``virtual'' field transmitted by the aperture complementary to the obstruction $f_A(x,0,z)$, and the field transmitted behind the obstruction $f_O(x,y,z)$. The plots are generated with $\vartheta_0 = \pi/32 \sim 6^\circ$, $k a = 5$ and $x_0 = 0 =y_0$. 
 For these values the geometrically predicted minimum reconstruction distance is equal to $z_0 = 2^{1/2} a/\tan \vartheta_0 \sim 72/k$. At $k z = 100$ the intensities of the fields $f(x,y,z)$ and $f_O(x,y,z)$ appear already very similar.}
\end{figure}
%
%
\\
From this figure it is evident the ``self-healing mechanism'' in action: during propagation from $z=0$, where the obstacle is located, to $k z = 100$  the field practically recovers its original intensity distribution. However, it should be noticed that 
due to the finite transverse extent of the obstruction, 
 the intensity $I_O(x,0,z) = \abs{f_O(x,0,z)}^2$ changes considerably during propagation only within the region  $\abs{kx} \lesssim 30$.  Conversely, it  remains constantly close to  $I(x,0,z) = \abs{f(x,0,z)}^2$ in the complementary region $\abs{kx} \gtrsim 30$. 
This phenomenon appears more clearly if one displays in the same figure the three intensity distributions at different $z$, as shown in  Fig. \ref{fig6} next page.

%
%
\begin{figure}[h]
\centerline{\includegraphics[clip=false,width=.8\columnwidth,trim = 0 50 0 50]{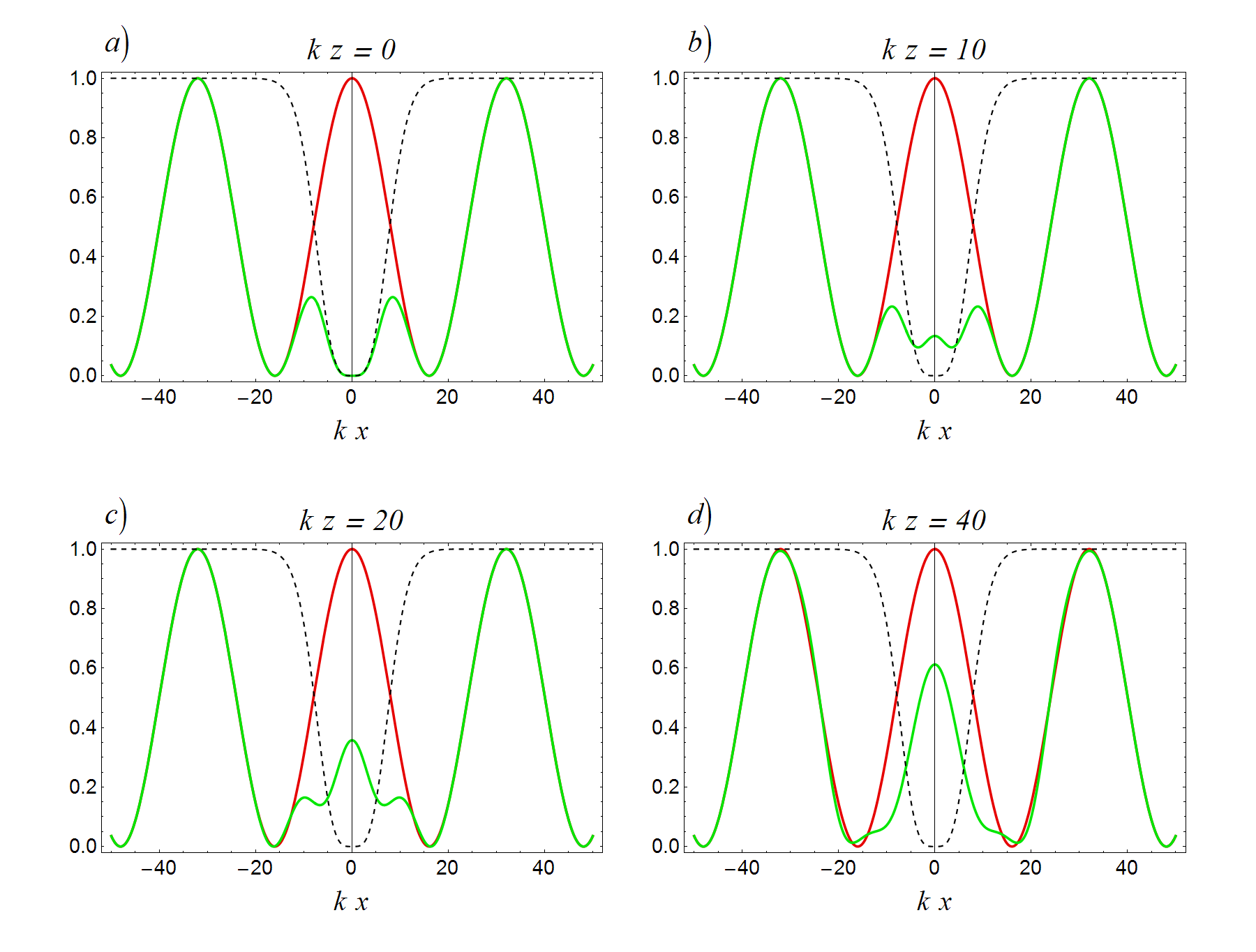}}
\caption{\label{fig6}
Intensity distributions $I(x,y,z) = \abs{f(x,y,z)}^2$ (red lines) and $I_O(x,y,z) = \abs{f_O(x,y,z)}^2$ (green lines) evaluated at $y=0$ and \emph{a}) $z =0$,  \emph{b}) $kz =10$, \emph{c}) $kz =20$ and \emph{d}) $kz =40$. The black dashed curves represent the intensity transmission function $\abs{t_O(x,y)}^2$ evaluated at $y=0$. The values of the parameters $A_0,\vartheta_0,a, x_0$ and $y_0$ are the same as in Fig. \ref{fig4}. It is evident how at \emph{any} $z \geq 0$ it has $I_O(x,0,z) \simeq I(x,0,z)$ for $ k \abs{x} \gtrsim 30 = 6 k a$.}
\end{figure}
%
%
This means that for all practical purposes Eq. \eqref{p40} represents a far too much restrictive constraint. What one really needs is simply to satisfy \eqref{p40} on the $xy$-plane in the neighborhood of the propagation axis $z$. This statement may be formalized as follows.  Consider again an obstruction whose orthogonal projection on the $xy$-plane occupies the region $O$,  and let $E$ be an arbitrary area in the $xy$-plane \emph{strictly} contained within  $O$, namely $E \subset O$. For example, $E$ can be the region confined by the inner circle of radius $b$ in  Fig. \ref{fig1}. Then, as necessary condition for self-haling, we require that
the amplitude $f_O(x,y,z)$ of the obstructed beam is proportional to the amplitude $f(x,y,z)$ of the unperturbed beam \emph{only}
 within  $E$:
\begin{align}\label{k100}
 f_O(x,y,z ) \Big|_{(x,y) \in E} \approx \lambda_0 f(x,y,z) \Big|_{(x,y) \in E} \qquad \forall z \geq z_0.
\end{align}

 From a mathematical point of view, Eq. \eqref{k100} makes much more sense than Eq. \eqref{p40}. In fact, according to the theory of \emph{angular spectrum representation} of a beam  \cite{MandelBook}, the field configuration at $z =0$ completely determines the field distribution at $z>0$. Therefore, if at a certain distance $z$ Eq. \eqref{p40} were satisfied upon \emph{all} the $xy$-plane, then it should be also valid at $z=0$. But the latter statement is clearly false because at $z=0$ one has, by definition,
\begin{align}\label{k110}
 f_O(x,y,0 ) = t_O(x,y) f(x,y,z) \neq f(x,y,z).
\end{align}
Thus, we have shown that the origin of the apparent paradox \eqref{q10} resides in the desideratum of satisfying both equations \eqref{p10} and \eqref{p40} over all the $xy$-plane. This is also the reason why the similarity defined in  equation (1) of Ref. \cite{Chu14}, fails to furnish a quantitative description of self-healing: The double integral defining the scalar product \eqref{k10} extends upon the whole $xy$-plane and this erases the $z$-dependence.

To circumvent this difficulty, in this work we propose to define the  scalar product in the space of functions $L_2(E)$ as
\begin{align}\label{k120}
\left( f, g \right) = \int_{ E } f^*(x,y,z) g(x,y,z)  \, dx dy, 
\end{align}
where the  integration is now restricted to the domain $E$. With this definition, the scalar product $\left( f, g \right) $ naturally becomes a function of $z$. Of course the definition \eqref{k120} is to some extent arbitrary in that the choice of the integration domain $E$ is partially discretionary (the only constraint is to be entirely contained within $O$). However, it is useful to remind here that the concepts themselves of ``self-healing'' and ``minimum reconstruction distance'' suffer from the same kind of arbitrariness. In other words, since both  Eqs. \eqref{p10} and \eqref{p40} are impossible to satisfy over the \emph{whole} $xy$-plane, one is forced to chose \emph{where} these equations should be satisfied. A reasonable choice is to take $E$ as the region bounded by the inner circle of radius $b \leq a$ in Fig. \ref{fig1}. However, different symmetries in the problem may dictate different choices, as we will explicitly show later in two examples.

Motivated by the introduction of the scalar product \eqref{k120} and recalling that the distance $d(f,g)$ between two functions in $L_2(E)$ can be defined as $d(f,g) = \left\| f -g \right\|$, where $\left\| f  \right\| = (f,f)^{1/2}$ \cite{KolmogorovBook}, we found it convenient to introduce as $z$-dependent witness  of self-healing, the  function 
\begin{align}\label{k125}
w(z) = 1 - d_r, 
\end{align}
 where the \emph{relative difference} $d_r$ is defined as  

\begin{align}\label{k130}
d_r = & \;\frac{\left\|f - f_O \right\|}{\left\|  f \right\| + \left\|  f_O \right\|}  \nonumber \\
= & \; \frac{\left(f_A , f_A \right)^{1/2}}{\left(f , f \right)^{1/2} +  \left[\left(f , f \right)-2 \operatorname{Re}\left(f , f_A \right) + \left(f_A , f_A \right) \right]^{1/2}} .
\end{align}
%
The limiting values of this witness function $w(z)$ are simply evaluated as follows. For a \emph{totally opaque} obstacle,  at $z=0$ it must be $f_O = 0$ for all points $(x,y) \in E$. Therefore, from the definition \eqref{k130} it follows that $w(0) = 0$. Vice versa, \emph{if} for $z \geq z_0$ it has $f_O \approx \lambda_0 f$ for  $(x,y) \in E$, then 
\begin{align}\label{k140}
w(z) \approx 1 - \left(\frac{1-\lambda_0}{1 + \lambda_0} \right)^{1/2} \to 1 \qquad \text{for} \qquad \lambda_0 \to 1.
\end{align}
Therefore, we have 
\begin{align}\label{k150}
0 \leq w(z) \leq 1.
\end{align}

In the remainder of this section we will test the effectiveness of our witness function by using it to asses the self-healing ability of: \emph{1.}  the two-plane-wave ``beam'' described above, and \emph{2.}  a Gaussian beam.

\subsubsection{Two-plane-wave beam}

\begin{figure}[h]
\centerline{\includegraphics[clip=false,width=.45\columnwidth,trim = 0 50 0 10]{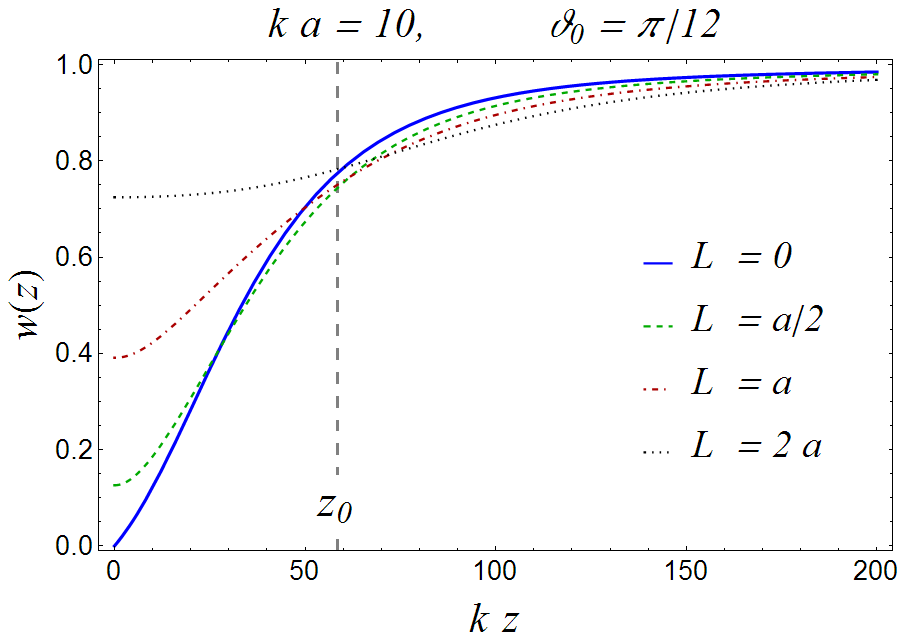}}
\caption{\label{fig7}
Plot of the witness function $w(z)$ given in Eq. \eqref{k125} evaluated for the two-plane-wave field \eqref{k50} and the Gaussian obstruction \eqref{k60} of width $k a =10$ and centered at $x_0=0=y_0$. The aperture of the ``beam'' is $\vartheta_0 = \pi/12$ and $z_0  \approx 58.5/k $ denotes the minimum reconstruction distance calculates as 
$z_0 = R(a)/\tan \theta $, where $R(a) = a \left[\ln \left(6 + 2^{5/2} \right) \right]^{1/2}$ denotes the half-width at half maximum of the intensity transmission function $\abs{t_O(x,0)}^2$.
It is evident that when $L$ increases, the dependence of $w(z)$ from $z$ becomes weaker and weaker. The continuous blue line represents the asymptotic value of $w(z)$ achieved for $L/a \to 0$.}
\end{figure}

The two-plane-wave field \eqref{k50} does not depend upon the variable $y$ and has essentially a Cartesian geometry. Hence,  we choose for $E$ a square of side $L$ centered at $(x=0,y=0)$, namely $E := \left\{(x,y)\in \mathbb{R}^2 : \abs{x} \leq L \wedge \abs{y} \leq L \right\}$. In this case the witness function $w(z)$ can be calculated analytically. However, the result of this calculation is very cumbersome and for the sake of clarity it will not be reported here. In Fig. \ref{fig7}  we display $w(z)$ as a function of $z$ for different sizes of the square region $E$.
\\
%
%
%
%
\\
Since a soft-edge Gaussian obstacle has not a sharp boundary, we can shrink the region $E$ to the single point $x=0=y$ and obtain the asymptotic value  for the witness function represented by the continuous blue line  in Fig. \ref{fig7} and denoted with $\bar{w}(z)$.

\begin{figure}[h]
\centerline{\includegraphics[clip=false,width=1\columnwidth,trim = 0 70 0 70]{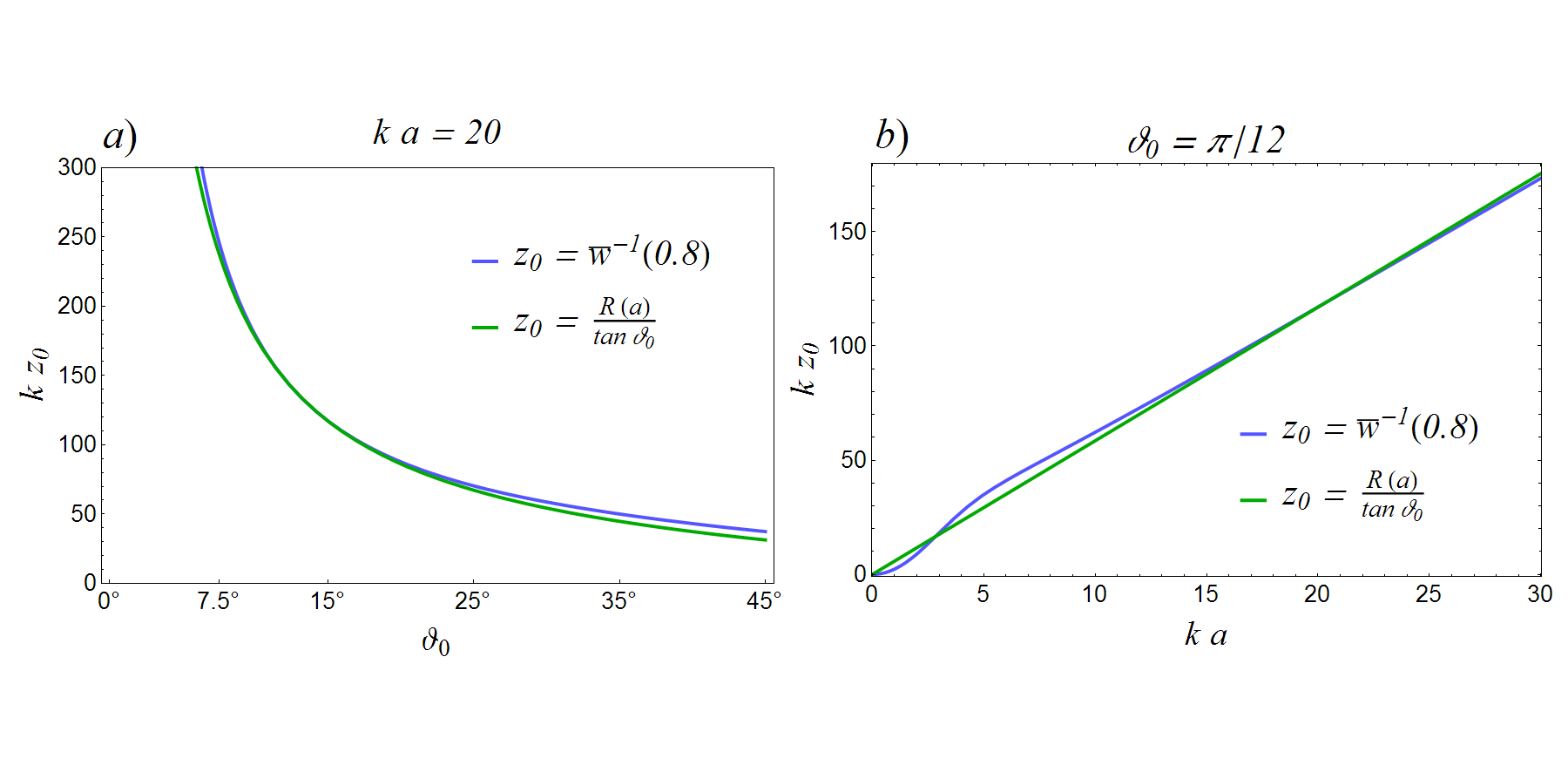}}
\caption{\label{fig8}
Plots of $\bar{w}^{-1}(\Delta)$ in Eq. \eqref{k160} as functions of either \emph{a}) the aperture $\vartheta_0$ and \emph{b}) the width of the transmission function  $a$. \emph{a}) Plot of $\bar{w}^{-1}(\Delta)$ (blue line) as a function of $\vartheta_0$, with $\Delta = 0.8$ and $k a =20$. For comparison, the plot of the geometrical-optics value $z_0 = R(a)/\tan \theta $ (green line), is shown.
 \emph{b}) Plot of $\bar{w}^{-1}(\Delta)$ (blue line) as a function of $a$, with $\Delta = 0.8$ and $\vartheta_0 = \pi/12$. For comparison, the plot of the geometrical-optics value $z_0 = R(a)/\tan \theta $ (green line), is shown. }
\end{figure}

The geometrical-optics value for the minimum reconstruction distance given in the literature, is obtained by calculating the point on the $z$-axis, namely the point at $x=0=y$, where the intensity of the obstructed field begin to raise from zero. Therefore, we can use the  asymptotic form  $\bar{w}(z)$, which is indeed calculated ``on-axis'', to estimate $z_0$ as follows. First, we choose a ``threshold'' value, say $\bar{w}=\Delta$, for the witness function. Then, we consider the beam as reconstructed only for those distances $z$ from the obstruction such that $\bar{w}(z) \geq \Delta$. By (numerically) inverting this relation we find the   minimum reconstruction distance $z_0$ as
\begin{align}\label{k160}
z_0 = \bar{w}^{-1}(\Delta).
\end{align}
The plots of $\bar{w}^{-1}(\Delta)$ as functions of either $\vartheta_0$ and $a$ are presented in Fig. \ref{fig8} where they are compared with the geometrical-optics values.\\
%
%
%
%
\\
In both cases the agreement between $\bar{w}^{-1}(\Delta)$ and the geometrical-optics value of $z_0$ is excellent.

\subsubsection{Gaussian beam}

\begin{figure}[h!]
\centerline{\includegraphics[clip=false,width=1\columnwidth,trim = 0 50 0 50]{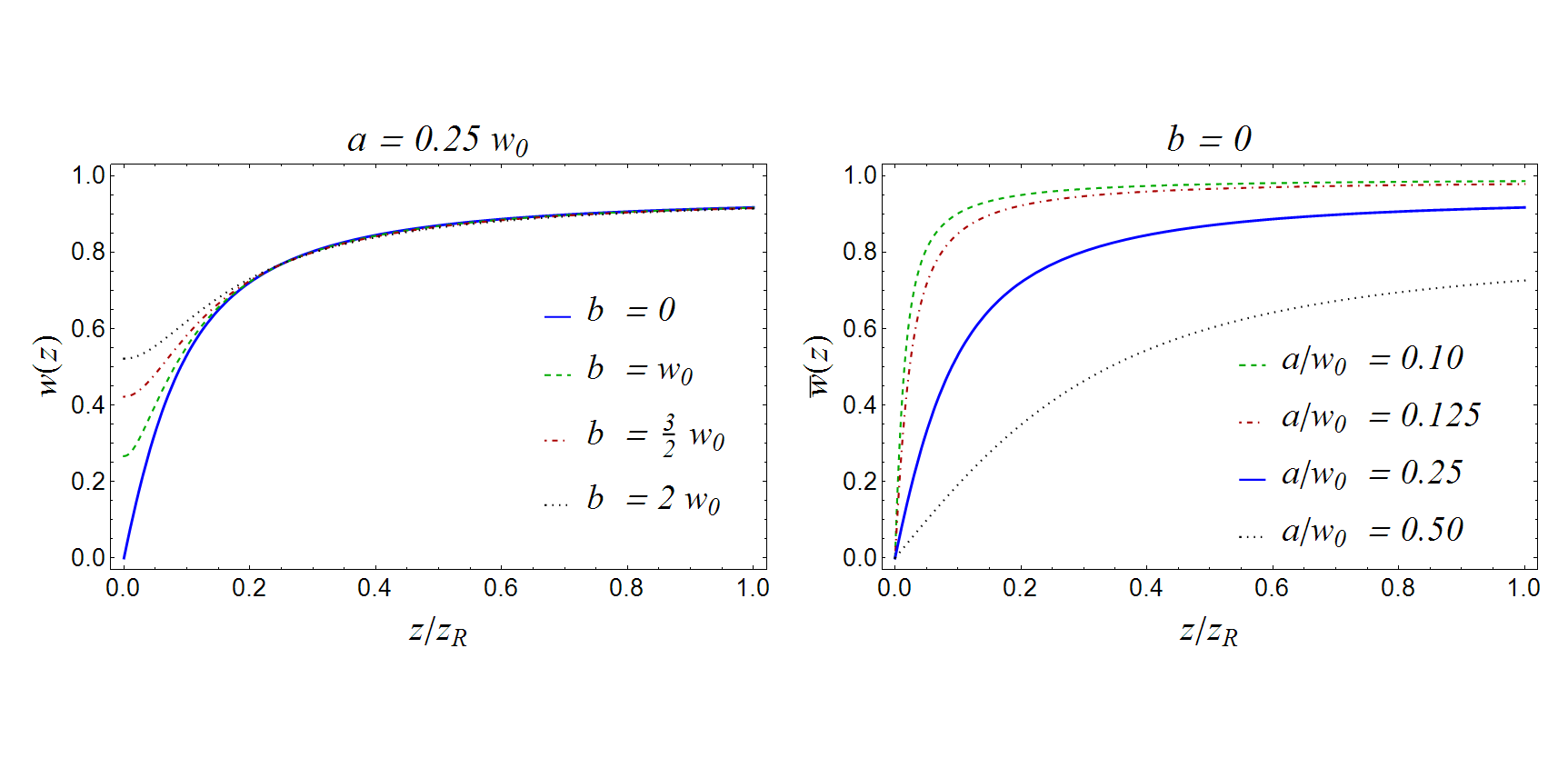}}
\caption{\label{fig9}
\emph{a}) Plots of the witness function $w(z)$ given in Eq. \eqref{k125}, evaluated for the Gaussian field \eqref{z120} of waist $w_0$ and different radii $b$ of the integration region $E$. 
It is evident that when  $b$ increases, the dependence of $w(z)$ from $z$ becomes weaker. The continuous blue line represents the asymptotic value  $\bar{w}(z)$ achieved for $b \to 0$.
\emph{b}) Plots of the asymptotic witness function $\bar{w}(z)$  \eqref{k170} for several values of the width $a$ of the soft-edge Gaussian obstruction. When the  obstruction shrink to zero, then  $\bar{w}(z) \to 1$, as expected from physical considerations.
}
\end{figure}

\begin{figure}[h!]
\centerline{\includegraphics[clip=false,width=1\columnwidth,trim = 0 0 0 0]{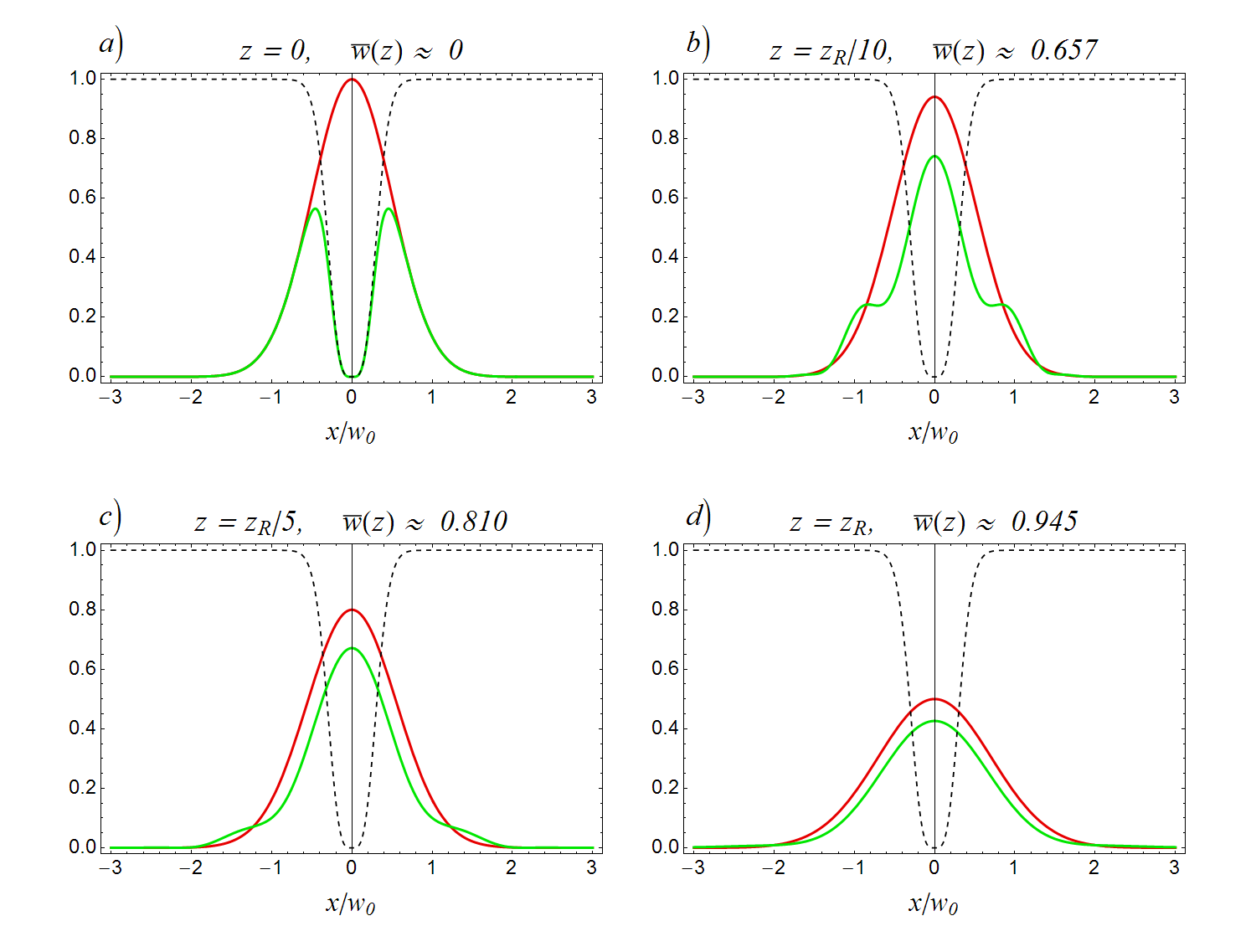}}
\caption{\label{fig10}
Intensity distributions $I(x,y,z) = \abs{g(x,y,z)}^2$ (red lines) and $I_O(x,y,z) = \abs{g_O(x,y,z)}^2$ (green lines) evaluated at $y=0$ and \emph{a}) $z =0$,  \emph{b}) $kz =10$, \emph{c}) $kz =20$ and \emph{d}) $kz =40$. The black dashed curves represent the intensity transmission function $\abs{t_O(x,y)}^2$ evaluated at $y=0$. Here $\theta_0=\pi/12$ and $a/w_0=0.2$.}
\end{figure}

The Gaussian beam studied in Sec. \ref{Gauss}, whose field is given by Eq. \eqref{z120}, possess a cylindrical symmetry about the propagation axis $z$. Therefore, now we choose for $E$ the disk of radius $b \leq a$ depicted in Fig. \ref{fig1}. Also in this case  $w(z)$ can be calculated analytically. In Fig. \ref{fig9}\emph{a})  we display $w(z)$ as a function of $z$ for different values of $b$. When $b$ goes to zero, we obtain the simple asymptotic form
\begin{align}\label{k170}
\bar{w}(\zeta) = 1 - \frac{\left( 1 + \zeta^2\right)^{1/2}}{\frac{\zeta}{2 \alpha^2} + \left[ 
1 + \frac{\zeta^2}{4 \alpha^4} \left( 1 + 2 \alpha^2\right)^2
\right]^{1/2}} ,
\end{align}
where $\zeta = z/z_R$ and $\alpha = a/w_0$. It is interesting to notice that this function is ``universal'' in the sense that it does not depend explicitly on the angular spread $\theta_0$ of the Gaussian beam. 
The plots of $\bar{w}(z)$ as function of  $z/z_R$ for different values of $a/w_0$, are presented in \ref{fig9}\emph{b}). It is evident that when the size $a$ of the obstruction goes to zero, the witness function tends to the unity. 
%
%
%
%
\\
In order to see the practical significance of $\bar{w}(z)$, in  Fig. \ref{fig10} next page we plot four sections  of the unperturbed Gaussian beam \eqref{z120} at different distances $z$ from the obstacle, and we compare it with the obstructed beam profile. The witness function $\bar{w}(z)$ evidently provide for a quantitative estimation of the similarity between these two fields.
\\
%
%

\section{Comparison between a Gaussian beam and a Bessel beam}

\begin{figure}[h!]
\centerline{\includegraphics[scale=3,clip=false,width=1\columnwidth,trim = 0 0 0 0]{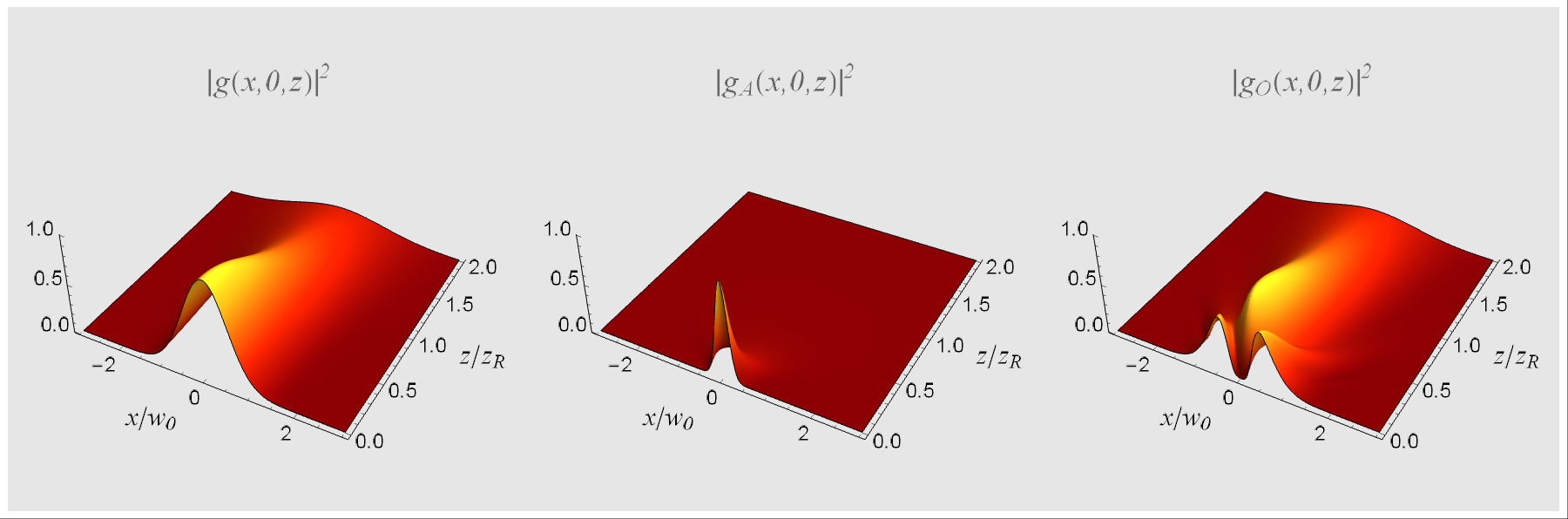}}
 \caption{ \label{fig110}
The plots are generated for a \emph{Gaussian beam} with angular aperture  $\theta_0 = 2/(k w_0) = \pi/32 \sim 6^\circ$ and a soft-edge Gaussian obstruction with full width $a/w_0 = 0.2$ and $x_0 = 0 =y_0$. 
}
\end{figure}

\begin{figure}[h!]
\centerline{\includegraphics[scale=3,clip=false,width=1\columnwidth,trim = 0 0 0 0]{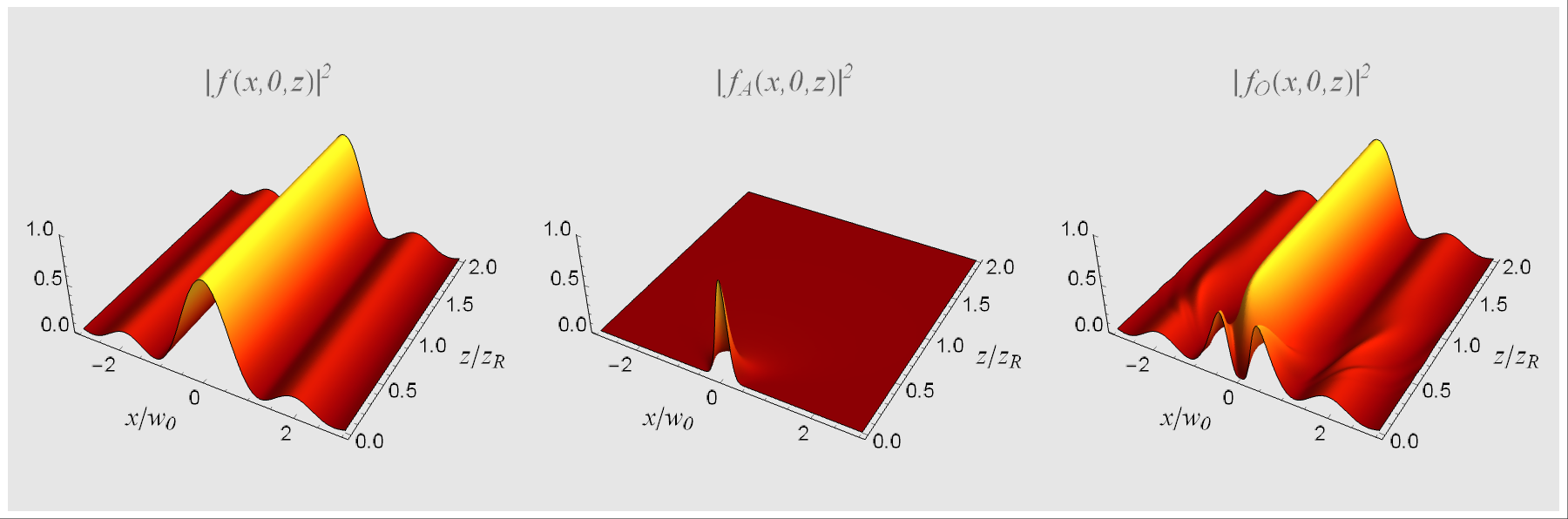}}
 \caption{ \label{fig120}
The plots are generated for a \emph{Bessel beam} with cone aperture  $\vartheta_0 = \pi/32 \sim 6^\circ$ and a soft-edge Gaussian obstruction with full width $a/w_0 = 0.2$ and $x_0 = 0 =y_0$, where $w_0 = 2/(k \vartheta_0)$. 
}
\end{figure}

In this last section we compare a Gaussian with a Bessel beam both transmitted behind a soft-edge Gaussian obstruction and propagating in the \emph{paraxial} regime. For comparison, we take the cone aperture $\vartheta_0$ of the Bessel beam equal to the angular spread $\theta_0$ of the Gaussian beam. In Figs. \ref{fig110} and  \ref{fig120} we plot the intensity distributions (absolute value squared) evaluated at $y=0$, of (left to right): the incident field, the ``virtual'' field transmitted by the aperture complementary to the obstruction, and the field transmitted behind the obstacle.
%
%
%
%
%
%
%
%
%
Next, we plot the asymptotic witness functions $\bar{w}_G(z)$ and $\bar{w}_B(z)$ for the Gaussian and the Bessel beam, respectively. The witness function is defined by the values of the intensity on the $z$-axis according to the formula
\begin{align}\label{m10}
\bar{w}(z) = 1 - \frac{\sqrt{I_A(z)}}{\sqrt{I(z)}+\sqrt{I_O(z)}} ,
\end{align}
where $I_{\#}(z) = \abs{f_{\#}(0,0,z)}^2$.
%
%
\begin{figure}[h!]
\centerline{\includegraphics[scale=3,clip=false,width=.6\columnwidth,trim = 0 0 0 0]{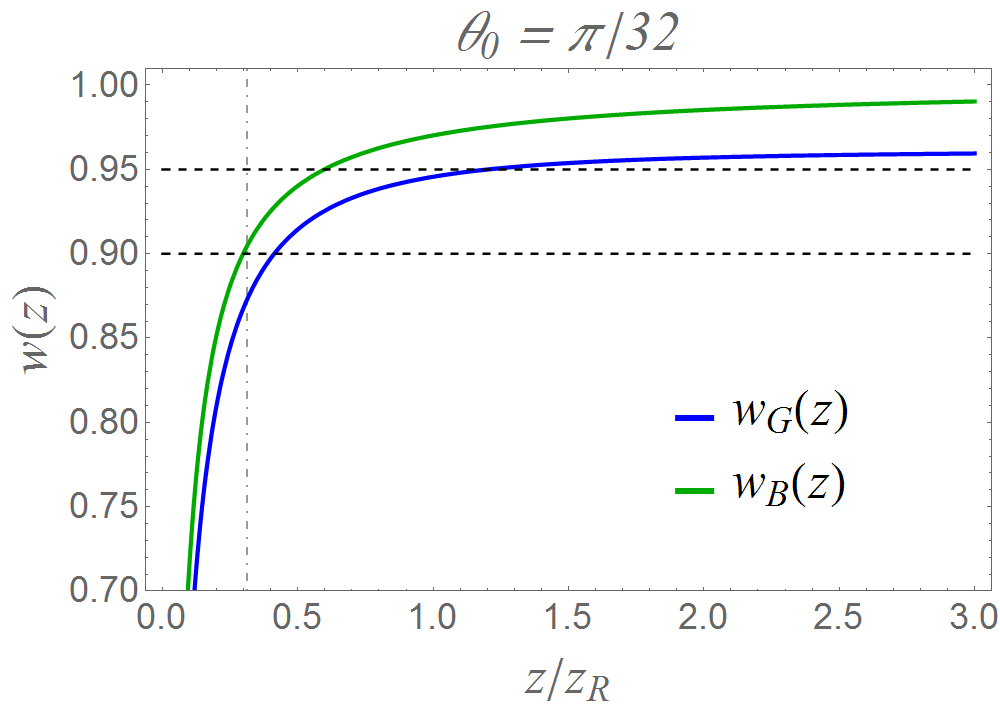}}
 \caption{ \label{fig30}
The plots are generated for  \emph{Bessel}  and  \emph{Gauss} beams with  $\theta_0 =\vartheta_0 = \pi/32 \sim 6^\circ$ and a soft-edge Gaussian obstruction with full width $a/w_0 = 0.2$ and $x_0 = 0 =y_0$, where $w_0 = 2/(k \theta_0)$. 
}
\end{figure}
%
%
In Fig. \ref{fig30} one can see that $\bar{w}_B(z) > \bar{w}_G(z)$ always, thus indicating a better self-healing property. The vertical line represents the geometrical-optics prediction for the minimum self-reconstruction distance $z_0/z_R \approx 0.312 $, with
\begin{align}\label{m20}
\bar{w}_G(z_0) \approx 0.87, \qquad \bar{w}_B(z_0) \approx 0.90.
\end{align}
The two horizontal lines represent the ``fidelity'' values $0.9$ and $0.95$ such that
\begin{align}\label{m30}
z_{0B} & \; = \bar{w}_B^{-1}(0.9) \approx 0.298,  & \;  z_{0B} & \; =   \bar{w}_B^{-1}(0.95) \approx 0.597, \\
z_{0G} & \;= \bar{w}_G^{-1}(0.9) \approx 0.413,  & \; z_{0G}  & \; =  \bar{w}_G^{-1}(0.95) \approx 1.20.
\end{align}
These results show that the Gauss beam reconstruct itself after the Bessel beam. Moreover, there is a limit to the reconstruction ability of a Gauss beam, because
\begin{align}\label{m40}
\lim_{z/z_R \to \infty}\bar{w}_G(z) = \frac{1}{1 + \frac{a^2}{w_0^2}}, \qquad \lim_{z/z_R \to \infty}\bar{w}_B(z) = 1,
\end{align}
so the fidelity for a Gaussian beam cannot reach the value $1$. This is understandable because the Gauss beam has a finite energy which is partially lost because of the obstruction.

\section{Conclusions}\label{Conc}

In this work we studied and compared the self-healing properties of  Gaussian, Bessel, Bessel-Gauss beams and of a pair of plane waves. We found a novel ``universal'' definition for the minimum reconstruction distance based on the angular spectrum distribution of a beam. Numerous examples are given in the text and illustrated with explanatory figures.

\appendix

\section{Notation}\label{Not}

Three-dimensional vectors in either real and Fourier space are denoted with \emph{Latin} letters: $\vec{r} = x \hat{x} + y \hat{y} + z \hat{z}$, $\vec{k}= k_x \hat{x} + k_y \hat{y} + k_z \hat{z}$. Two-dimensional vectors in either real and Fourier space are denoted with \emph{Greek} letters: $\vec{\rho} = x \hat{x} + y \hat{y}$, $\vec{\kappa}= k_x \hat{x} + k_y \hat{y}$. Cylindrical coordinates in real and Fourier spaces are denoted with $\left(\rho,\phi,z \right)$ and $\left(\kappa,\varphi, k_z \right)$, respectively, with $\rho^2 = x^2 + y^2$ and $\kappa^2 = k_x^2 + k_y^2$. 
All fields considered here are monochromatic with wave number $k$ and angular frequency $\omega = c \, k$.

The \emph{Fourier transform} of a function $g(x,y)$ of two independent variables $x$ and $y$ will be denoted either by $\mF[g](k_x,k_y)$ or by $G(k_x,k_y)$ and is defined by
\begin{align}\label{n10}
G(k_x,k_y) = \frac{1}{2 \pi} \iint\limits_{-\infty}^{\phantom{xx}\infty} g(x,y) \exp \left[- i \left( x k_x + y k_y\right) \right] \, dx dy. 
\end{align}
Similarly, the \emph{inverse Fourier transform} of a function $G(k_x,k_y)$ will be represented either by $\mF^{-1}[G](x,y)$ or by $g(x,y)$ and is defined as
\begin{align}\label{n20}
g(x,y) = \frac{1}{2 \pi} \iint\limits_{-\infty}^{\phantom{xx}\infty} G(k_x,k_y) \exp \left[i \left( x k_x + y k_y\right) \right] \, dk_x dk_y. 
\end{align}

The two-dimensional Dirac delta symbol $\delta(\vka - \vka')$ stands for
\begin{align}\label{n30}
\delta(\vka - \vka') = \delta(k_x-k_x') \delta(k_y-k_y'),
\end{align}
and is defined as
\begin{align}\label{n40}
\delta(\vka - \vka') = \frac{1}{(2 \pi)^2} \iint\limits_{-\infty}^{\phantom{xx}\infty}  \exp \left[i x \left( k_x -k_x'\right) \right] \exp \left[i y \left( k_y -k_y'\right) \right] \, dx dy. 
\end{align}

\section{The Chu\&Wen proposal}\label{CW}

In the article \cite{Chu14}, Chu\&Wen  defined the  scalar product in the space of functions $L_2(\mathbb{R}^2)$ as
\begin{align}\label{k10}
 \langle f, g \rangle = \iint\limits_{-\infty}^{\phantom{xx}\infty} f^*(x,y,z) g(x,y,z)  \, dx dy. 
\end{align}
Later, in their equation (1), they proposed the following measure of the similarity between two given functions $f$ and $g$:
\begin{align}\label{k20}
\text{Similarity} \; \sim \; \cos \left( f \cdot g \right) = \frac{\langle f, g \rangle}{\left\|f\right\| \left\|g\right\|} ,
\end{align}
where the \emph{norm} of a function is defined as $\left\|f\right\| = \langle f, f \rangle^{1/2}$. This notation may seems not very appropriate because the right side of Eq. \eqref{k20} can be a complex number. A more suitable definition yielding to a non-negative number is simply 
\begin{align}\label{k30}
\text{Similarity} \; \sim \;  \frac{\abs{\langle f, g \rangle}}{\left\|f\right\| \left\|g\right\|} ,
\end{align}
which coincides with the standard definition of \emph{fidelity of quantum pure states} in quantum mechanics \cite{NielsenBook}.

Using the Parseval's theorem, Chu\&Wen were able to show that the similarity defined as in Eq. \eqref{k20} cannot depend upon the propagation distance $z$ (in their derivation Chu\&Wen implicitly assumed that the considered angular spectrum did not contain evanescent waves). Therefore, they adopted a new description of similarity, given in their equation (11), defined as
\begin{align}\label{k40}
\text{Similarity} \; \sim \; \cos \left( \abs{f} \cdot \abs{g} \right) = \frac{\langle \abs{f}, \abs{g} \rangle}{\left\|f\right\| \left\|g\right\|} .
\end{align}
With this definition, the similarity between $f$ and $g$ becomes a function of the propagation distance $z$.
However, if one applies Eq. \eqref{k40} to pure (namely, non Gaussian) Bessel beams, this similarity function becomes $z$-independent. Therefore, also the definition  \eqref{k40} in some circumstances may be not fully satisfactory. 
%

%
%
%
%
%
%
%

%

\end{document}